\documentclass[a4paper,twocolumn,showpacs,preprintnumbers,showkeys]{revtex4}
\usepackage{fullpage}
\usepackage{graphicx}
\usepackage{subfigure}
\usepackage{epsfig}
\usepackage{dcolumn}
\usepackage{bm}
\usepackage{amsmath,amssymb}

\begin{document}

\title{ Single Particle Stochastic Heat Engine }

\author{Shubhashis Rana$^1$, P. S. Pal$^1$, Arnab Saha$^2$, A. M. Jayannavar$^1$}

\email{shubho@iopb.res.in, priyo@iopb.res.in, sahaarn@gmail.com, jayan@iopb.res.in}

\affiliation{Institute of Physics, Sachivalaya Marg, Bhubaneswar - 751005, India$^1$ \\ Institut f$\ddot{o}$r Theoretische Physik II, Weiche Materie,
Heinrich-Heine-Universit$\ddot{a}$t D$\ddot{u}$sseldorf,
40225 D$\ddot{u}$sseldorf, Germany$^2$}






\begin{abstract}
We have performed an extensive  analysis of a single particle stochastic heat engine constructed 
by manipulating a  Brownian particle in a time dependent harmonic potential. The cycle consists
 of two isothermal steps at different temperatures and two adiabatic steps similar to that of a Carnot
 engine. The engine shows qualitative differences in inertial and overdamped regimes.
All the thermodynamic quantities, including  efficiency, exhibit strong fluctuations in a time periodic steady state.
 The fluctuations of stochastic efficiency dominate over the mean values even
 in the quasistatic regime. Interestingly, our system acts as an engine provided the temperature difference between the two 
reservoirs is greater than a finite critical value which in turn depends on the cycle time and other system parameters. This is supported by
our analytical results carried out in the quasistatic regime. Our system works more reliably as an engine
for large cycle times. By studying various model systems we observe that the operational characteristics are model dependent.
Our results clearly rules out any universal relation between efficiency at maximum power and temperature of the baths.
 We have also verified fluctuation relations for heat engines in time periodic steady state.
  
\end{abstract}

\maketitle

\newcommand{\nwc}{\newcommand}
\nwc{\vs}{\vspace}
\nwc{\hs}{\hspace}
\nwc{\la}{\langle}
\nwc{\ra}{\rangle}
\nwc{\lw}{\linewidth}
\nwc{\nn}{\nonumber}

\nwc{\pd}[2]{\frac{\partial #1}{\partial #2}}
\nwc{\zprl}[3]{Phys. Rev. Lett. ~{\bf #1},~#2~(#3)}
\nwc{\zpre}[3]{Phys. Rev. E ~{\bf #1},~#2~(#3)}
\nwc{\zpra}[3]{Phys. Rev. A ~{\bf #1},~#2~(#3)}
\nwc{\zjsm}[3]{J. Stat. Mech. ~{\bf #1},~#2~(#3)}
\nwc{\zepjb}[3]{Eur. Phys. J. B ~{\bf #1},~#2~(#3)}
\nwc{\zrmp}[3]{Rev. Mod. Phys. ~{\bf #1},~#2~(#3)}
\nwc{\zepl}[3]{Europhys. Lett. ~{\bf #1},~#2~(#3)}
\nwc{\zjsp}[3]{J. Stat. Phys. ~{\bf #1},~#2~(#3)}
\nwc{\zptps}[3]{Prog. Theor. Phys. Suppl. ~{\bf #1},~#2~(#3)}
\nwc{\zpt}[3]{Physics Today ~{\bf #1},~#2~(#3)}
\nwc{\zap}[3]{Adv. Phys. ~{\bf #1},~#2~(#3)}
\nwc{\zjpcm}[3]{J. Phys. Condens. Matter ~{\bf #1},~#2~(#3)}
\nwc{\zjpa}[3]{J. Phys. A  ~{\bf #1},~#2~(#3)}

\section{Introduction}

Thermodynamic heat engines convert heat into useful work. They work cyclically  between two thermal
 reservoirs kept at different temperatures $T_l$ and $T_h$ ($T_h > T_l$). The Second law of thermodynamics restricts their 
efficiency to the Carnot limit \cite{car24},  $\eta_{C}=1-\frac{T_l}{T_h}$. However, this efficiency can only be achieved in the quasistatic limit where transitions between 
thermodynamic states occur infinitesimally slowly and hence the power output vanishes. Curzon and Ahlborn (C-A) \cite{cur75}
showed that  for finite time endoreversible heat engines, efficiency at maximum power is given by
  $\eta_{CA}=1-\sqrt{\frac{T_l}{T_h}}$. As yet there is no consensus  on this result (\cite{sei12,sch08,bro05,esp10,tu14}).

With the advances in nano-technology, a few-micrometer-sized Stirling heat engine has been experimentally realized \cite{nphy12}.This microscopic 
heat engine operates in conditions where typical changes in their energies are of the order of the thermal
energy per degree of freedom \cite{rit05}. An appropriate theoretical framework to deal with these systems  has been developed 
during the past decades within the context of stochastic thermodynamics \cite{sek-book,sek98,dan05,sai07,jop08}.
This formalism of stochastic energetics provides a method to calculate  work, heat and entropy even for a single particle  along a microscopic
 trajectory. One can obtain  average quantities after averaging over respective ensembles. The averaged thermodynamic quantities, 
 work and entropy, obey Second law. Using this formulation various single particle heat engines 
have been studied in the literature \cite{sch08,eng13,tu13,hol14,esp14}. Fluctuation relations for heat engines (FRHE) \cite{sin11,lah12,cam14} operating in a 
time periodic steady state (TPSS) have recently been obtained \cite{lah12}. FRHE are in the form of equality and Carnot's inequality  for efficiency
$\eta_c$ follows as a direct consequence of this theorem.

In the present work we have studied in detail a simple model for a stochastic heat engine described by Langevin 
equation. Both underdamped and overdamped regimes are explored and qualitative differences are pointed out. We
emphasize on fluctuations of thermodynamic variables including the engine efficiency. We show that 
fluctuations dominate the mean values even in quasistatic regime. Therefore in such situations one needs to study the full probability 
distribution of the physical variable for the proper analysis of the system. 

In section \ref{mod}, we describe the model of our system and the protocol. In section \ref{uql}, we obtain analytical results for 
relevant average thermodynamic  quantities in the quasistatic regime for the underdamped case. In  section
\ref{unl}, engine with finite time cycle in the inertial regime is studied numerically, in detail. The system driven by 
time asymmetric cycles and various other  model systems are also explored. We have verified FRHE in this section.
Sections \ref{oql} and \ref{onl} are devoted to the  analytical  and numerical studies of the system in 
the overdamped limit. Finally, we conclude in section \ref{con}. Each section is  self contained.

\section{The Model}
\label{mod}
\begin{figure}

\includegraphics[width=7.5cm]{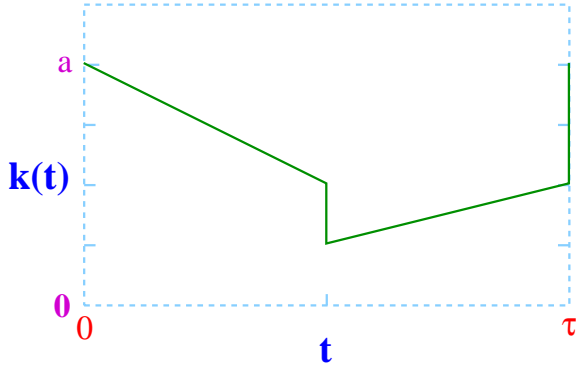}

\caption{(Color online) Stochastic heat engine in a harmonic potential: the time dependence of our periodically driven stiffness
 constant (protocol) k(t) for the full cycle ($0\leq t\leq \tau$). }
\label{fig-pro}
\end{figure}

The single particle stochastic heat engine consists of a Brownian particle having position $x$ and
velocity  $v$ at time $t$, confined in a  one dimensional harmonic trap. The stiffness of the trap $k(t)$ varies periodically in time as shown 
in Fig.(\ref{fig-pro}). For the underdamped case, the equation of motion for the particle is given by \cite{risk-book,coffey-book}

\begin{equation}
  m \dot v=-\gamma v-k(t)x+\sqrt{\gamma T}\xi(t)  .
\label{u-lan}
\end{equation}
In overdamped limit the equation reduces to

\begin{equation}
\gamma\dot x=-k(t)x+\sqrt{\gamma T}\xi(t) .
\label{u-lan1}
\end{equation}

In our further analysis, we  set mass of the particle m, the Boltzmann constant $k_B$ and the frictional coefficient $\gamma$  to be unity.
 $T$ is the temperature of the thermal bath. All  physical parameters are made dimensionless.
 The noise is Gaussian with zero mean, $\la \xi(t)\ra=0 $  and is  delta correlated, $\la \xi(t_1)\xi(t_2)\ra=2\delta(t_1 -t_2)$. 
The internal energies of the particle  in the underdamped and the overdamped limit are given by $u(x,v)=\frac{1}{2}k(t)x^2+\frac{1}{2}mv^2$
and  $u(x)=\frac{1}{2}k(t)x^2$, respectively.

\begin{figure}

\includegraphics[width=7.5cm]{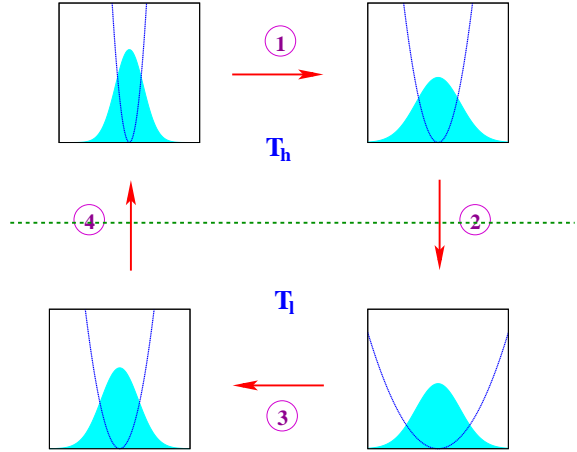}

\caption{(Color online) Schematic representation for a cyclic process of stochastic heat engine operating between two reservoirs kept at  temperatures $T_h$ and $T_l$.
The cycle consists of two isothermal  steps and two adiabatic steps according to the time varying protocol k(t). 
The blue line denotes a one dimensional potential V(x,t) and the filled region denote the corresponding steady state distribution.}
\label{fig-car}
\end{figure}

Operation of the system consists of four steps - two isotherms and two adiabatics. In the first step, the system undergoes an
 isothermal expansion, during which it is connected to a hot bath at temperature $T_h$ and the stiffness constant 
is varied linearly with time  as
\begin{equation}
 k(t)=a\left(1-\frac{t}{\tau}\right)=k_1(t)
\label{u-lan2}
\end{equation}
for $0<t<\tau/2$. Here  $\tau$ is the period of the cycle and $a$
 is the initial value of the stiffness constant. In the second step, the potential undergoes an instantaneous expansion
 (adiabatic) by decreasing the stiffness constant from $a/2$ to $a/4$. As the process is instantaneous the distribution before and after 
expansion will not change and heat absorption will be zero. In the third step, the system is
  connected to a cold bath with lower temperature $T_l$ and isothermal compression of the trap is carried out
  by changing the stiffness  as
\begin{equation}
k(t)=a\frac{t}{2\tau}=k_2(t) 
\label{u-lan3}
\end{equation}
for $\tau/2<t<\tau$. In the last step, we carry out instantaneous adiabatic compression by varying the stiffness
 constant from $a/2$ to $a$  and simultaneously connecting the system  to the hot bath.
This cycle is then repeated . The time dependence of the protocol is given in Fig.(\ref{fig-pro}) and a schematic representation of the 
system within a cycle at its various stages is depicted in Fig.(\ref{fig-car}).  

The described protocol differs from those used in earlier studies. In the experimental set up \cite{nphy12}  two adiabatic steps are absent. Work 
optimized protocol is used by Schmiedl and  Seifert \cite{sch08} whereas the protocol based on the concept of shortcut 
to adiabaticity is used by Tu \cite{tu13}. However, their emphasis is on the possible correlation between efficiency at
maximum power and C-A bound. Our main motivation, namely to study the fluctuation of physical quantities, is different from earlier studies as mentioned in the introduction.

\section{Underdamped quasistatic limit}
\label{uql}
In this section, we  analytically calculate the average thermodynamic quantities of our model system in the quasistatic limit.
In this limit, the duration of the protocol is much larger than all the  relevant time scales, including the relaxation time.
Hence as protocol is changed, the system immediately adjusts to the equilibrium state corresponding to the value
of protocol at that instant. First, we  calculate the average work done on the particle in all the four steps of a
 cycle and the heat absorbed by it in the first isothermal step. Finally, we calculate efficiency in the quasistatic limit.  

In the first isothermal process, average work done on the particle is the same as the free 
energy change $(\Delta F_h)$ before and after the expansion, i.e.,
\begin{equation}
W_1=\Delta F_h=\frac{T_h}{2}\ln\frac{k_1(\tau/2)}{k_1(\tau)}=\frac{T_h}{2}\ln\frac{1}{2}   .
\end{equation}

At $t=\tau/2$, the system  is in equilibrium  with the bath at $T_h$  with 
stiffness constant $a/2$. The second step being instantaneous,  no heat will be dissipated and the phase space distribution remains 
unaltered. Correspondingly the average  work done on the particle is equal to the change in its internal energy:

\begin{eqnarray}
 W_2&&=N_1\int_{-\infty}^{\infty}dx dv\left(\frac{a}{4}-\frac{a}{2}\right)\frac{x^2}{2}e^{-\frac{a x^2}{4 T_h}-\frac{v^2}{2T_h}}\nn\\
&&=-\frac{T_h}{4},
\end{eqnarray}
where $N_1=\frac{1}{2\pi T_h}\sqrt{\frac{a }{2}}$, is the normalization constant. Similarly in the third step 
(i.e., isothermal compression step) the average work done on the particle in the quasistatic limit is
\begin{equation}
W_3=\Delta F_l=\frac{T_l}{2}\ln\frac{k_2(\tau)}{k_2(\tau/2)}=\frac{T_l}{2}\ln2.
\end{equation}
The average work done in the last step (i.e., second adiabatic step) is given  as
 \begin{eqnarray}
 W_4&&=N_2\int_{-\infty}^{\infty}dx dv(a-\frac{a}{2})\frac{x^2}{2}e^{-\frac{a x^2}{4 T_l}-\frac{v^2}{2T_l}}\nn\\
&&=\frac{T_l}{2},
\end{eqnarray}
with $N_2=\frac{1}{2\pi T_l}\sqrt{\frac{a }{2}}$. Hence, the average total work done in the full cycle of 
the heat engine in the quasistatic process is 
\begin{eqnarray}
 W_{tot}&&=W_1 +W_2+W_3 +W_4\nn\\
&&=\frac{T_h}{2}\ln\frac{1}{2}-\frac{T_h}{4}
                               +\frac{T_l}{2}\ln2+\frac{T_l}{2}.
\label{eq-w}
\end{eqnarray}
To obtain the heat absorption in the first step (i.e., isothermal expansion), we  calculate the average change of 
internal energy and use the First law. During this process, the particle stays in contact
 with hot bath at temperature $T_h$. However, it is to be noted that at time $t=0^{-}$, the system was in contact with 
low temperature bath at $T_l$, whereas at $t=0^{+}$  the system is in contact with hot bath at $T_h$. Thus the system
has to relax into new equilibrium after sudden change in temperature. The time taken for this relaxation
process is assumed to be negligible compared to the cycle time $\tau$. This relaxation leads to an additional heat flow 
which  accounts for the change in the internal energy during the relaxation process.   One can readily obtain
the internal energy at $t=0^{+}$ as $3T_l/2$ while after the relaxation it is $T_h$.
 Hence, the average internal energy change in the first step is

\begin{equation}
 \Delta U_1 =T_h-\frac{3T_l}{2}.
\end{equation}

Now using the First law, the average heat absorption from the hot bath for the first step is
\begin{equation}
 -Q_1=\Delta U_1 -W_1=T_h-\frac{3T_l}{2}-\frac{T_h}{2}\ln\frac{1}{2}.
\label{eq-q} 
\end{equation}
Hence efficiency of the engine for the underdamped case in the quasistatic limit is given by

\begin{eqnarray}
  \bar{\eta}_{q} &&=\frac{-W_{tot}}{-Q_1}=-\frac{\frac{T_h}{2}\ln\frac{1}{2}-\frac{T_h}{4}  +\frac{T_l}{2}\ln2+\frac{T_l}{2}}
                                {T_h-\frac{3T_l}{2}-\frac{T_h}{2}\ln\frac{1}{2}}\nn\\
&&=-\frac{T_h\ln\frac{1}{2}-\frac{T_h}{2}  +T_l\ln2+T_l}
                                {2T_h-3T_l-T_h\ln\frac{1}{2}}.
\label{uq-eff}
\end{eqnarray}

Here we would like to emphasize that $ \bar{\eta}$ is defined ignoring fluctuations and the subscript q denotes the quasistatic limit.
 We will show later that fluctuations
play an important role even in the quasistatic regime. Work done during the cycle $w$ and heat absorbed during the first step
$q_1$ are fluctuating quantities. Stochastic efficiency is defined as $\eta=\frac{w}{q_1}$ \cite{esp14} and hence its average
 $\la\eta\ra=\la\frac{w}{q_1}\ra$  is not the same as $\bar{\eta}=\frac{\la w\ra}{\la q_1\ra}$ which is given in Eq.(\ref{uq-eff}) for quasistatic limit. 
This will be discussed in detail in subsequent sections. In our notation, the thermodynamic quantities are denoted by capital
letters only for quasistatic limit, whereas, small letters are used to denote those quantities for finite time cycles.

According to our convention negative work done on the system implies extraction of work; while, negative heat
 means that heat enters into the system. It is important to note from Eq.(\ref{eq-q}) that in quasistatic limit 
 heat flows from the bath to the system provided 
\begin{eqnarray}
 &&2T_h-3T_l-T_h\ln\frac{1}{2}\ge 0\nn\\
&& \Rightarrow \frac{T_l}{T_h}\le \frac{2+\ln 2}{3}=0.898
\end{eqnarray}
and similarly from Eq.(\ref{eq-w}) work can be extracted from the system if
\begin{eqnarray}
&& T_h\ln\frac{1}{2}-\frac{T_h}{2}  +T_l\ln2+T_l\le 0\nn\\
&& \Rightarrow\frac{T_l}{T_h}\le \frac{0.5+\ln 2}{1+\ln 2}=0.705.
\end{eqnarray}
 Therefore, in quasistatic regime  our model system operates in three different modes of operation depending on 
the ratio of the  temperatures of the thermal baths. First, when
 $0 < \frac{T_l}{T_h}\le 0.705$ is maintained, work can be extracted
 and heat is absorbed from hot bath and it acts as an engine. Second, when
 $0.705 \le \frac{T_l}{T_h}\le 0.898$ is set, heat is absorbed 
from the bath but we cannot extract work. And finally when we have $\frac{T_l}{T_h}\ge 0.898$ 
neither heat is absorbed nor the work is extracted. In this case work done on the system heats up the hot bath.
Therefore, there is a particular regime in parameter space where the system act as an engine. This is in contrast to the 
Carnot engine which works for arbitrary temperature difference between two baths. The above mentioned condition
 is only valid in the quasistatic limit. For finite time cycle the operational condition for heat engine
depends on cycle time apart from $T_h$ and $T_l$, which will be shown in our simulation. Our exact expression 
of $W_{tot}$ and $Q_1$ are in complete  agreement with our numerical results in the quasistatic limit. Thus these analytical
calculations act as a check for our numerical simulation.

\section{finite cycle time engine in inertial regime}
\label{unl}
 For finite-cycle-time we study our system numerically.
When the Langevin system is driven periodically it is known that after initial transients, the system will settle 
down to a TPSS. The joint probability distribution $P_{ss}(x,v,t)$ of position and velocity of the particle 
is periodic in time, i.e., $P_{ss}(x,v,t)=P_{ss}(x,v,t+\tau)$.

For numerical simulations  we evolve our system with a time periodic protocol (as shown in Fig.\ref{fig-pro}).
 We have used Heun's method for integrating the basic Langevin equation \cite{man00} with time step $dt=0.0002$.
 We make sure that the system is in the TPSS by going beyond the initial transient regime. We then 
consider at least $10^5$ cycles of operations and physical quantities are averaged over all these cycles. 
 For rest of the paper  we keep m, a, $\gamma$ fixed at $m=1.0$, $a=5.0$, $\gamma =1.0$.

We now make use of the concepts of stochastic energetics \cite{sek-book,sek98,dan05,sai07,jop08} to calculate work, heat and internal energy for a given
trajectory. The thermodynamic  work done on the particle during first part of the cycle, in each computational step $dt$, is given by

\begin{equation}
 dw_1(t_i)=\pd{u_1(t_i)}{k_1(t_i)}\dot k_1(t_i) dt.
\end{equation}
with $u_1(t_i)=\frac{1}{2}k_1(t_i)x^2(t_i) + \frac{1}{2}v^2(t_i)$ and $t_i=i.dt $. Now, $w_1=\sum _{i=0}^{N}dw_1(t_i)$
 where $N=\frac{\tau}{2dt}$. The internal energy is a thermodynamic state function and hence its change during the 
isothermal process is given by  $du_1=\frac{1}{2}k_1(\tau/2)x^2(\tau/2) + \frac{1}{2}v^2(\tau/2)-\frac{1}{2}k_1(0)x^2(0)-\frac{1}{2}v^2(0)$.
 The heat absorption by the bath is $q_1=w_1-du_1$ using the First law of thermodynamics.  
 The second step which is adiabatic is instantaneous and hence the particle does not get any chance to evolve. Thus work done is only 
instantaneous change in internal energy, i.e., $w_{2}=\frac{1}{2}[k_2(\tau/2)-k_1(\tau/2)]x^2(\tau/2)$.  
Similarly, for step three, work done is given by 
\begin{equation}
 dw_3(t_i)=\pd{u_2(t_i)}{k_2(t_i)}\dot k_2(t_i) dt
\end{equation}
and $w_3=\sum _{i=N}^{2N}dw_3(t_i)    $; internal energy change
 $du_2=\frac{1}{2}k_2(\tau)x^2(\tau)+\frac{1}{2}v^2(\tau)-\frac{1}{2}k_2(\tau/2)x^2(\tau/2)-\frac{1}{2}v^2(\tau/2)$ ;
and heat delivered to the cold bath is $q_2=w_3-du_2$. For the last adiabatic process, 
  work done on the particle is  $w_4=\frac{1}{2}[k_1(0)-k_2(\tau)]x^2(\tau)$.  
The total work done on the system in a cycle is $w=w_1+w_2+w_3+w_4$.
It should be noted that each $w_i$ (i=1,2,3,4) is a fluctuating quantity and their values depend on a particular phase 
space trajectory.

\begin{figure}

\includegraphics[width=7.5cm]{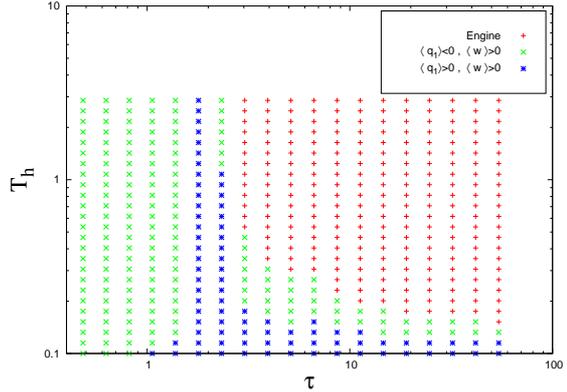}

\caption{(Color online) Phase diagram for different $T_h$ and $\tau$ but for fixed $T_l=0.1$. 
}
\label{uf-phase}
\end{figure}
In Fig.(\ref{uf-phase}), we have shown the phase diagram of the operation of our system. Here we have 
 varied $T_h$ and cycle time $\tau$ keeping $T_l$ fixed at 0.1. There are three distinct regimes. The system acts as 
an engine when $\la w\ra < 0$ and $\la q_1\ra < 0$. The angular bracket $\la.\ra$ indicates average over several
realizations. In the other two regimes the system ceases to work as a heat engine
altogether( $\la w\ra > 0$). For $\la w\ra>0$ we have two distinct domains with   $\la q_1\ra < 0$ and  $\la q_1\ra > 0$. 
The latter  implies work is done on the system which heats up the hot bath. In the large cycle time limit numerical 
results are consistent with our analytical predictions made in last section. We re-emphasize that the system works 
as a heat engine provided there  is a minimal difference between $T_h$ and $T_l$ which depends on cycle time $\tau$ and
other physical parameters. From the Phase diagram it is apparent that, as we decrease $\tau$ for fixed $T_h$, there exists a lower bound below which
 the system does not perform as an engine, as it only consumes work.

\begin{figure}
\vspace{0.5cm}

\includegraphics[width=7.5cm]{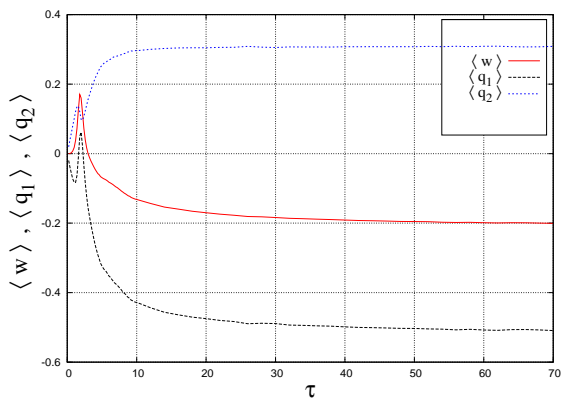}
\caption{(Color online) Variation of$ \la w \ra$, $\la q_1\ra$ and $\la q_2\ra$ with cycle time $\tau$.
}
\label{uf-work}
\end{figure}

In Fig.(\ref{uf-work}), we have plotted $\la w\ra$, $\la q_1\ra$ and $\la q_2\ra$ with respect to cycle time $\tau$. 
We have fixed $T_h=0.5$ and $T_l=0.1$ for all subsequent figures.
Starting from zero,  $\la w\ra$ initially increases and reaches a peak value. Then it starts decreasing
and finally saturates to a negative value ( -0.214),  which is close  to our theoretical result (from Eq.(\ref{eq-w})).
The work can be extracted in the region where it becomes negative. As we increase cycle time, $\la q_1\ra$ changes dramatically. It has a positive region sandwiched between
 two negative regions. When $\la q_1\ra>0$ heat is released to the hot bath 
 while work is done on the particle. In the quasistatic limit it saturates at the theoretical value -0.523(from Eq.(\ref{eq-q})).
 In contrast to $\la q_1\ra$, $\la q_2\ra$ is always positive, i.e., heat is always released to the cold bath. 
Internal energy being a state function, $\la \Delta u\ra$  is zero over a cycle in TPSS and hence $\la w\ra=\la q_1\ra+\la q_2\ra$.
Using the saturation value of $\la w\ra$ and $\la q_1\ra$ we immediately get $\la q_2\ra$  to be 
equal to 0.310 which is close to our numerical result.

We now study the nature of stochastic efficiency $\eta$ and engine power $p=-\frac{w}{\tau}$ as a function of cycle time.  
 The engine is in TPSS where probability
 distributions of system variables are periodic in time. However, for a given 
realization of a cycle, state of the system (position and velocity) does not come back to its initial state. Thus 
for each cycle thermodynamic quantities will depend on the particular microscopic trajectory and hence 
$w$, $q_1$, $q_2$, $\eta$ and $p$  are all fluctuating quantities from cycle to cycle. The average efficiency is 
defined as $\la \eta \ra=\la\frac{w}{q_1}\ra$. Due to fluctuation in $w$ and $q_1$, it is to be noted that 
$\la\eta\ra=\la\frac{w}{q_1}\ra\ne \frac{\la w\ra}{\la q_1\ra}=\bar{\eta}$. Fluctuation theorems \cite{sin11,lah12,cam14} put stringent condition on
$\frac{\la w\ra}{\la q_1\ra}$ which is bounded by the Carnot efficiency i.e., $\frac{\la w\ra}{\la q_1\ra}\le 1-\frac{T_l}{T_h}$.
However, no such bound exist for $\la\eta\ra$ \cite{cam14}.

\begin{figure}
\vspace{0.5cm}

\includegraphics[width=7.5cm]{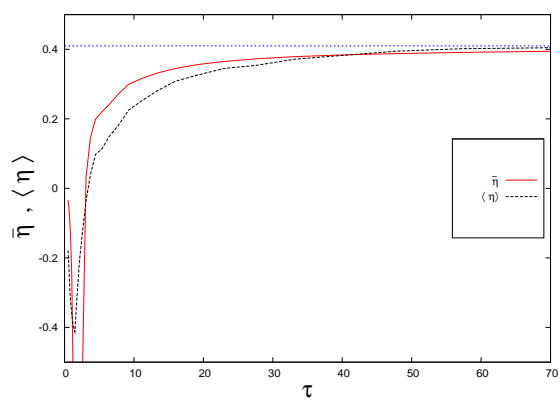}
\caption{(Color online) Variation of $\la \eta\ra$ and $\bar{\eta}$ with cycle time $\tau$.The doted blue line denotes the 
quasistatic limit for  $\bar{\eta}$.
}
\label{uf-eff}
\end{figure}

The First law  for any microscopic realization of cycle can be written as
\begin{equation}
 w=\Delta u + q_1+ q_2.
\label{first-law}
\end{equation}
The change in the internal energy $\Delta u$  is unbounded. It is zero only on the average. Similarly $q_1, q_2$ and $w$
take values in the range ($-\infty,\infty$) but are constrained by First law. Hence it is not surprising that $\eta$  
 can take values between $-\infty$ to $\infty$.

In Fig.(\ref{uf-eff}) we have plotted efficiencies $\la\eta\ra$ and $\bar{\eta}$ as a function of cycle time. Initially for small $\tau$,
our system doesn't work as an engine. Due to large dissipation work cannot be extracted ($\la w\ra \ge 0$). In this 
regime, efficiency is negative. On further increasing $\tau$, efficiency becomes positive and it monotonically 
increases. For large $\tau$, $\la\eta\ra$ and $\bar{\eta}$ saturate. The saturation value for $\bar{\eta}$ is 0.41 which
can be obtained analytically in quasistatic regime. In general  $\la\eta\ra \ne \bar{\eta}$.
We find both  $\la\eta\ra$ and $\bar{\eta}$ are less than the Carnot efficiency $\eta_c=0.8$.

\begin{figure}
\vspace{0.5cm}

\includegraphics[width=7.5cm]{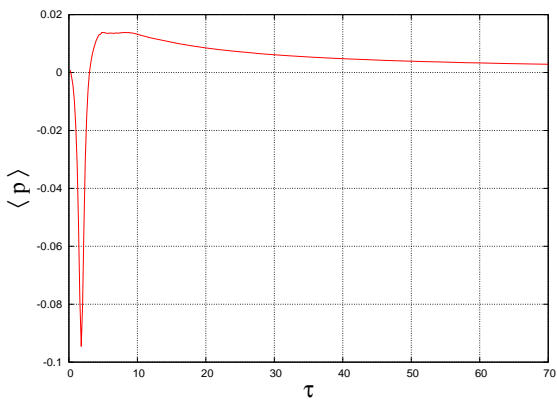}
\caption{(Color online) Variation of  average power $\la p\ra$ with  cycle time $\tau$.
}
\label{uf-pow}
\end{figure}

 In Fig.(\ref{uf-pow}), average power $\la p\ra$ is plotted as a function of $\tau$. There is a negative region for low cycle 
time. Beyond the critical value of $\tau\simeq 3.0$, power becomes positive and exhibits a peak and finally tends to zero in the large
$\tau$ limit. The efficiencies $\la\eta\ra$ and $\bar{\eta}$ at maximum power are given by 0.16 and 0.25 respectively. Both of these values are less than $\eta_{CA}=0.554$.

\begin{figure}

\vspace{0.5cm}

\includegraphics[width=7.5cm]{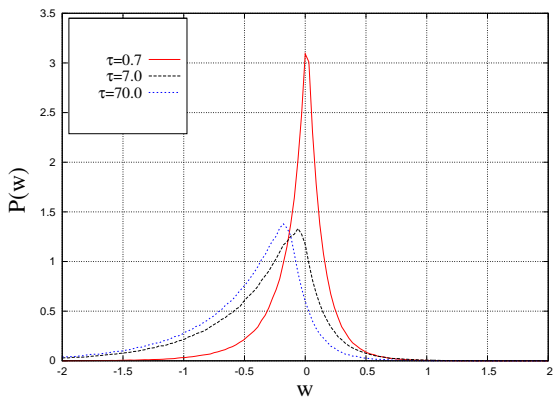}
\caption{(Color online) Distribution of $w$ for different cycle times ($\tau$=0.7, 7.0, 70.0).
}
\label{uf-wdist}
\end{figure}

\begin{figure}
\vspace{0.5cm}

\includegraphics[width=7.5cm]{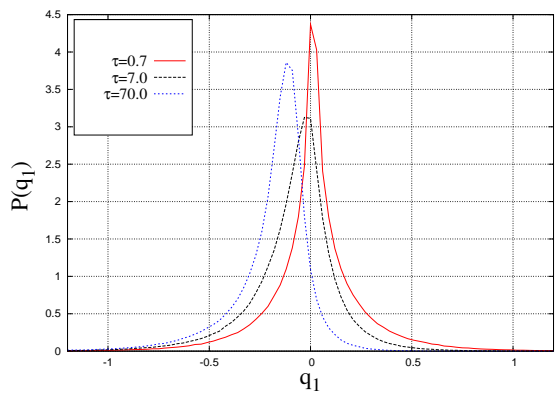}
\caption{(Color online) Distribution of $q_1$ for different cycle times.
}
\label{uf-Q1dist}
\end{figure}

\begin{figure}
\vspace{0.5cm}
\includegraphics[width=7.5cm]{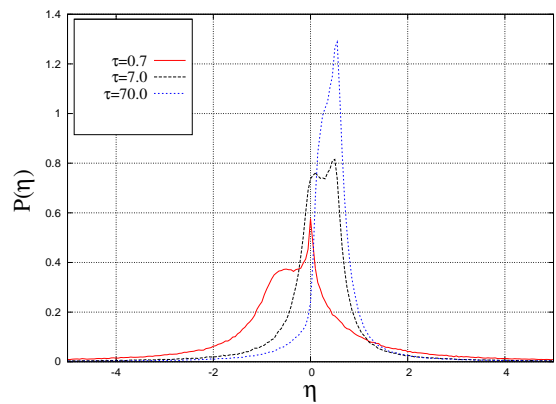}
\caption{(Color online) Distribution of $\eta$ for different cycle times.
}
\label{uf-effdist}
\end{figure}

As mentioned earlier,  physical quantities $q_1$, $w$ and $\eta$ are strongly fluctuating variables. To study
 these fluctuations we focus on probability distribution of these quantities $P(q_1)$, $P(w)$ and 
$P(\eta)$. In figs.(\ref{uf-wdist}), (\ref{uf-Q1dist}) and (\ref{uf-effdist}) we have plotted them for three different time periods. For $\tau=0.7$, distribution
of $w$ and $q_1$ are sharply peaked around zero with $\la w\ra=0.005$, $\la q_1\ra=-0.065$. As we increase the cycle
 time   $P(w)$ and $P(q_1) $ become broad, asymmetric and  shift towards negative
side. For large negative value of arguments the distributions exhibit long tail. For large positive values of $w$ and $q_1$ 
the distribution falls off exponentially or faster \cite{jar10}. The trajectory responsible for  positive values are atypical
and sometimes referred to as transient Second law violating trajectories \cite{rit03,wang02,sahoo11}. Strong fluctuations in heat and work persist even in
 the quasistatic limit($\tau=70$). These fluctuations in work are mainly attributed to two adiabatic processes, while 
fluctuations of $q_1$ result from relaxation process when the system, in contact with low temperature  bath, is
brought in direct contact with high temperature reservoir.


\begin{figure*}[ht]
\vspace{0.5cm}

\includegraphics[width=15cm]{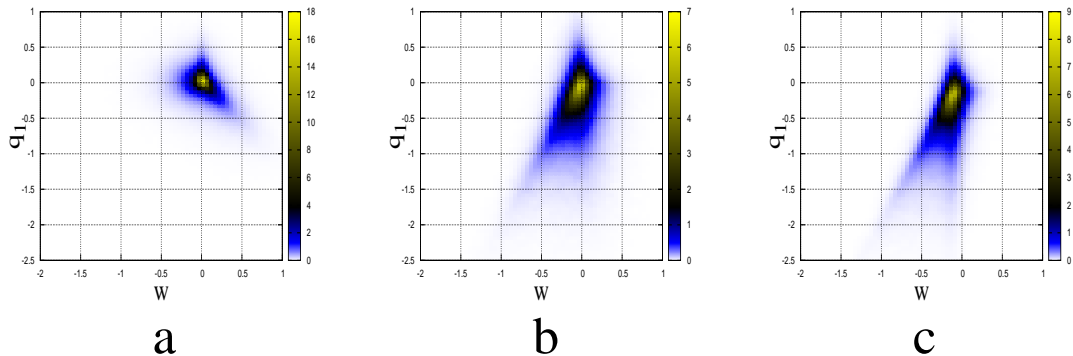}
\caption{(Color online) Joint distribution of $w$ and $q_1$ for different $\tau$. In a)  $\tau=0.7$, in b)  $\tau=7.0$, in c)  $\tau=70$.
}
\label{tau0.7}
\end{figure*}

For  $\tau=0.7$, $\la\eta\ra$ is negative (-0.26). The distribution $P(\eta)$ is asymmetric and 
there is a broad shoulder on the negative side. As we increase $\tau$, distribution shifts towards positive side.
 It is not surprising to see the finite weight for values 
$\eta < 0$ and $\eta > 1$ \cite{cam14}. As we increase the cycle time the standard deviation 
of $\eta$ ($\sigma_{\eta}$), becomes smaller. However, it remains larger compared to mean values. For example, 
$\la \eta\ra=0.161$ and corresponding $\sigma_{\eta}=1.32$ at $\tau=7.0$ and $\la \eta\ra=0.406$ 
whereas  $\sigma_{\eta}=1.11$ for $\tau=70.0$. We would like to emphasize that mean is dominated by fluctuations
even in the quasistatic regime. Any physical quantity with relative variance 
 larger than one, is  referred to as non-self
averaging quantity. For such cases mean ceases to be a good physical variable and one has to resort to the analysis
for full probability distribution. This is one of our main result. Non-self averaging quantities arises mainly 
in physics of quenched disordered systems.

Note that, $\eta$ becomes positive if both $w$ and $q_1$ are positive or both of 
them are negative. $\eta$ becomes negative when  $w$ and $q_1$ have opposite signs.
In order to have a better understanding of  our system we have plotted the joint distributions of $w$ and $q_1$
for different $\tau$ in Fig.(\ref{tau0.7}). For a given cycle, the system acts 
as an engine when both $w$ and $q_1$ are negative i.e, in the third quadrant of the plot.
Using our numerical results we have calculated the ratio of the total number of realizations falling in the third 
quadrant  to the total number of realizations. These fractions 
 for $\tau=0.7$, 7.0 and 70.0 are calculated to be 0.226, 0.583 and 0.858, respectively. It is clear 
from this that for large cycle times the reliability of the system working as an engine increases.
Though we observe that even in quasistatic regime there are realizations for which the system does not 
act as an engine. This is due to strong fluctuations in work and heat as discussed earlier.

\begin{figure*}[ht]
\vspace{0.5cm}

\includegraphics[width=15cm]{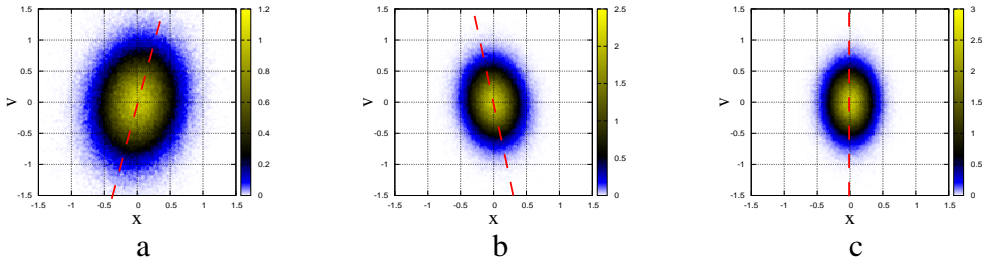}
\caption{(Color online) Initial phase space distribution at different cycle times $\tau$. In a)$\tau=0.7$, in b)  $\tau=7.0$, in c)$\tau=70$.
 The asymmetric position of red broken line along the major axis of the
elliptical Gaussian distribution for lower values of $\tau$ (=0.7 and 7.0)
indicates nonzero $\langle xv\rangle$. This correlation becomes zero for
larger $\tau$ (=70), where the position of the major axis also becomes
symmetric.}
\label{u-3d}
\end{figure*}

In TPSS the joint probability density $P_{ss}(x,v,t)$ is periodic in time:  $P_{ss}(x,v,t+\tau)$= $P_{ss}(x,v,t)$.
For simplicity we write $ P_{ss}(x,v,t)=e^{-\phi(x,v,t)}$. From the definition of stochastic entropy \cite{sei05,sei08,lah09} of the system $S_{sys}$,
the change in the system entropy for a trajectory over a cycle is given by $\Delta S_{sys}=\Delta \phi=\phi(x_2,v_2,\tau)-\phi(x_1,v_1,0)$
 where $(x_1,v_1)$ and $(x_2,v_2)$
 are the initial and final  phase space points for  a particular realization of the cycle.
To calculate $\Delta \phi$ we  evaluate $P_{ss}(x,v,0)$ at the initial point of the cycle which also coincide 
at the end point $t=\tau$. In Fig.(\ref{u-3d}) we have plotted joint phase space
distribution at TPSS for three different values of $\tau=0.7$,  $7.0$ and 70.0. We see that for $\tau=0.7$ and $\tau=7.0$
phase space distributions  are not symmetric and there exist strong correlation between $x$ and $v$ which was ignored in the 
earlier literature \cite{tu13}. Only in the large $\tau$ limit  the distribution becomes symmetric (see Fig.(\ref{u-3d}c)). The cross-correlation between position and velocity 
disappears and the distribution $P_{ss}(x,v)$ becomes uncorrelated Gaussian in the quasistatic limit. Due to correlation, the width of the distribution
become larger as we decrease cycle time $\tau$.

Recently, FRHE  in TPSS has been derived \cite{lah12}. It extends the total 
entropy production fluctuation theorem of Seifert \cite{sei05,sei08,arnab09} applied to heat engine. The total entropy production $\Delta S_{tot}$
over a cycle is a stochastic variable and in our present case is given by
\begin{equation}
 \Delta S_{tot}=\Delta \phi + \frac{q_1}{T_h} +\frac{q_2}{T_l}.
\end{equation}
Using the First law (Eq.\ref{first-law})
 \begin{equation}
 \Delta S_{tot}=\Delta \phi + \frac{q_1}{T_h} +\frac{w-q_1-\Delta u}{T_l}.
\end{equation}
The  Second law which is valid on average, can be stated as  $\la\Delta S_{tot}\ra\ge 0$. In TPSS, $\la\Delta u\ra = \la \Delta \phi\ra=0$,
which implies $\bar{\eta}=\frac{\la w\ra}{\la q_1\ra}\le 1-\frac{T_l}{T_h}=\eta_c$. Thus the Second law puts the constraint
on efficiency which is defined as  $\frac{\la w\ra}{\la q_1\ra}$.
It should be noted that this constraint is valid for any finite time cycle in TPSS, unlike the Carnot which is 
valid for macroscopic engines in the quasistatic regime. However, it does not put any constraint on the average efficiency 
($\la \frac{w}{q_1}\ra$). The fluctuation theorem for heat engine replaces the inequality relation of the 
Second law by the equality relation, namely \cite{lah12},

\begin{figure}
\vspace{0.5cm}

\includegraphics[width=7.5cm]{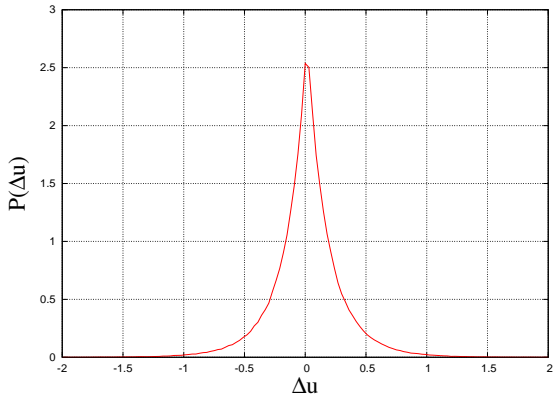}
\caption{(Color online)  Distribution of the internal energy change in one cycle for underdamped steady state for $\tau$=7.0.
}
\label{uf-u}
\end{figure}

\begin{figure}
\vspace{0.5cm}

\includegraphics[width=7.5cm]{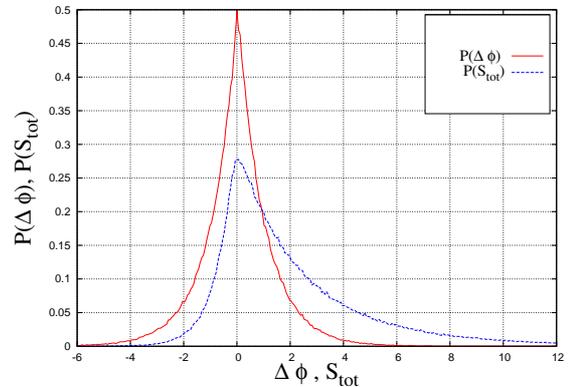}
\caption{(Color online) Distribution of system entropy change and total entropy production in one cycle for
 underdamped steady state for $\tau$=7.0}.
\label{uf-ent}
\end{figure}
\begin{equation}
\la e^{-\Delta S_{tot}}\ra =\la e^{-(\Delta \phi + \frac{q_1}{T_h} +\frac{w-q_1-\Delta u}{T_l})}\ra =1
\label{IFT}
\end{equation}
Eq.(\ref{IFT}) is FRHE in TPSS. By calculating all the relevant stochastic variables
$w$, $q_1$, $\Delta \phi$, $\Delta u$ over all trajectories  for finite $\tau$ we have verified  Eq.(\ref{IFT})
in the TPSS. We have obtained  the value to be 0.96 for $\tau=7.0$, which is well within our numerical accuracy.
We would like to emphasize that, in Eq.(\ref{IFT}), four stochastic variables appear in the exponent. Small changes in these 
values affect the exponential function by a large amount. Given this fact, our observed value of $\la e^{-\Delta S_{tot}}\ra$
is quite satisfactory. 

For the same parameter value $\tau=7.0$, in Fig.(\ref{uf-u}) we have plotted the probability distribution, 
$P(\Delta u)$, as a function of $\Delta u$. 
In Fig.(\ref{uf-ent}), we have plotted the probability distribution of change of system entropy $P(\Delta \phi)$
 and total entropy $P(\Delta S_{tot})$ as a function of their arguments. It is clear that as $u$
and $ \phi$ are state functions, $P(\Delta u)$ and $P(\Delta \phi)$ are symmetric with zero mean.
 However, the distribution  $P(\Delta S_{tot})$ is asymmetric  with a long tail for positive large $\Delta S_{tot}$.
There is also a finite weight towards negative $\Delta S_{tot}$. This contribution arises due to transient Second 
law violating periodic cycles \cite{rit03,sahoo11}. However,  $\la\Delta S_{tot}\ra$ remains positive as demanded by
the Second law.

\begin{figure}
\vspace{0.5cm}

\includegraphics[width=7.5cm]{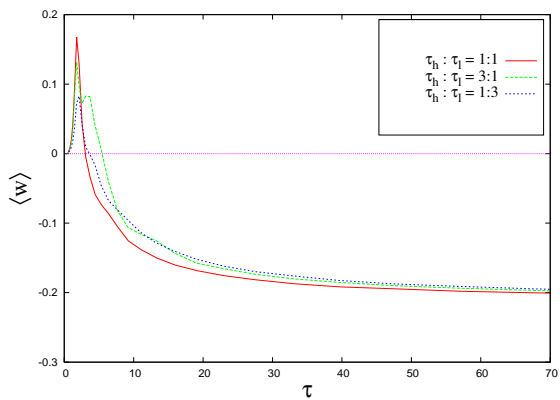}
\caption{(Color online) Variation of $\la w\ra$ vs $\tau$ for symmetric as well as asymmetrical cycles.
 Here $\tau_h$ and $\tau_l$ are contact times of the particle with hot and
cold bath respectively. Thus, $\tau=\tau_h+\tau_l =7.0 $ for our case.
 }
\label{uw-new}
\end{figure}

\begin{figure}
\vspace{0.5cm}

\includegraphics[width=7.5cm]{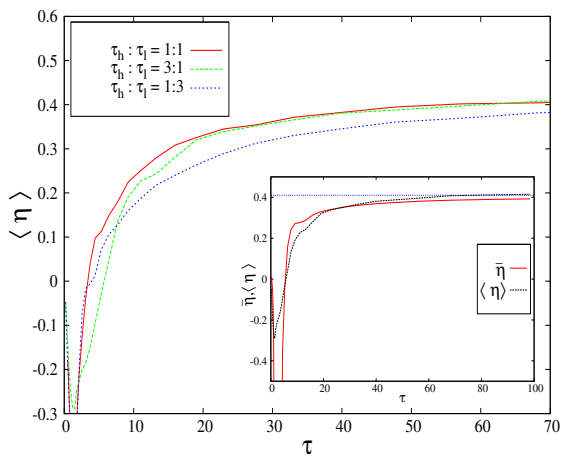}
\caption{(Color online) Variation of $\la \eta \ra$ vs $\tau$ for symmetric as well as asymmetrical cycles.
inset: comparision of $\la\eta\ra$ with $\bar{\eta}$ for $\tau_h:\tau_l=3:1$.}
\label{ueff-new}
\end{figure}

Till now we concentrated on symmetric cycle, i.e., equal  contact times of the system with hot and cold bath.
 Naturally, the question arises what will happen if the cycle is  time asymmetric. To the best of our knowledge this question has not 
been addressed in  earlier literature. If the contact time of one bath
 is  different from that of the  other, it can affect  work output, heat dissipation to each bath, power and efficiency.
However, in the quasistatic limit there should not be any effect of this asymmetry. This is clear from  Fig.(\ref{uw-new})
 that the average work, for three different asymmetric cycles, asymptotically approach each other in the quasistatic limit. In the non-quasistatic 
limit,  work extracted by the engine for asymmetric cycles is small compared to symmetric cycle.
 From Fig.(\ref{ueff-new}) it is seen that $\la\eta\ra$ is lower for asymmetrical cycles. The inset shows even in quasistatic limit
 $\la\eta\ra \ne \bar{\eta}$  for $\tau_h:\tau_l=3:1$. We have verified separately that asymmetry also decreases the power. Thus asymmetry in the cycle degrades the 
performance characteristics of the engine.

We now briefly compare the nature of power and efficiency of our system when the confining potential is different.
We have taken the confining potential $\frac{1}{2}k(t)x^n$ with n=2,4,6. For n=4,6 the confining potentials are 
 referred to as  hard potential. The equilibrium distributions for hard potentials are no longer Gaussian and hence in the quasistatic 
limit, the average work, heat dissipation etc., will be different from those for the harmonic potential.

\begin{figure}
\vspace{0.5cm}

\includegraphics[width=7.5cm]{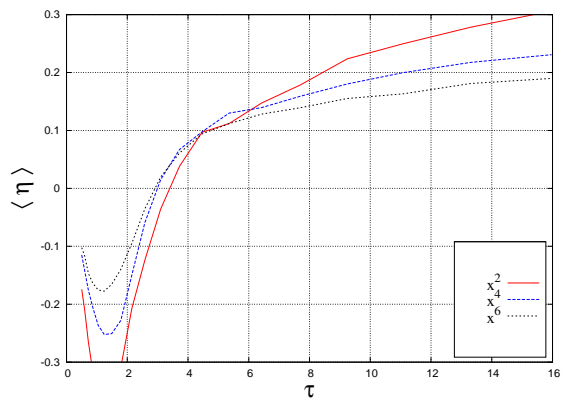}
\caption{(Color online) Variation of $\la \eta\ra$ with $\tau$ for different types of potentials.}
\label{uh-eff}
\end{figure}

\begin{figure}
\vspace{0.5cm}

\includegraphics[width=7.5cm]{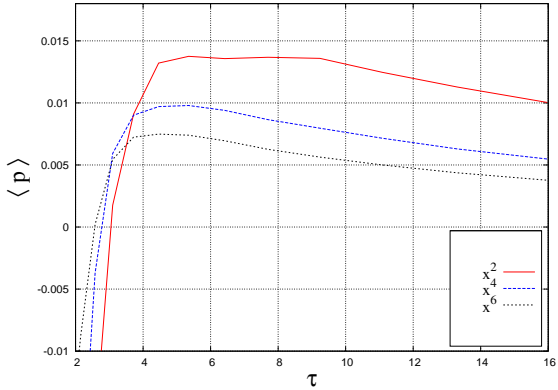}
\caption{(Color online) Variation of power $\la p\ra$ with  $\tau$ for different types of potentials.  }
\label{uh-pow}
\end{figure}

In Fig.(\ref{uh-eff}) and (\ref{uh-pow}) we have plotted $\la\eta\ra$ and $\la p\ra$ as a function of cycle time for
 different potentials.  Average efficiency $\la\eta\ra$ for large $\tau$ decreases as potential becomes harder 
and thereby degrading the performance. $\la\eta\ra$ saturates at the higher value of $\tau$ (not shown in the figure). 
From Fig.(\ref{uh-pow}) we  observe that harder the potential  smaller will be the critical
time $\tau$ above which the system acts as an engine. For large cycle time the power decreases  as the potential
becomes harder. However, we see clearly that there are three values of efficiencies $\la\eta\ra$ and $\bar{\eta}$
 at maximum power 0.16, 0.10, 0.08 and 0.25, 0.16, 0.13 for n=2,4,6 respectively.
It is apparent that the efficiency at maximum power is model dependent and decreases 
as the potential becomes harder.  Even the saturation value is different and it is lower for harder potential.
Clearly, these two figures indicate that operational characteristics of our system are model dependent. 
Thus we do not expect any universal relation involving only the average efficiency at maximum power and
temperatures of the reservoirs.
So far we have studied our system  in detail in the underdamped regime which is a general case. From now on we 
restrict to the overdamped regime and highlight some qualitative differences. 

\section{Overdamped quasistatic case}
\label{oql}
In the overdamped limit, dynamics of the system follows Langevin Eq.(\ref{u-lan1}), where inertial effects are ignored. This 
approximation is valid when the time steps of the observation are much larger than $m/\gamma$. The internal energy
of the system is given only in terms of potential energy. For this case equilibrium
distribution of a particle in a static harmonic potential is given by $P_{eq}(x)=N e^{-\frac{kx^2}{2k_B T}}$ from which one can
easily obtain the free energy. The analytical calculation for average thermodynamic quantities in quasistatic limit are similar to the 
underdamped case. The total average work done on the particle during the entire cycle is given by  
as
\begin{eqnarray}
 W_{tot}&&=\Delta F_h +W_2+\Delta F_l +W_4\nn\\
&&=\frac{T_h}{2}\ln\frac{1}{2}-\frac{T_h}{4}
                               +\frac{T_l}{2}\ln2+\frac{T_l}{2}.
\label{o-wtot}
\end{eqnarray}
Interestingly, the expression for $W_{tot}$ remains the same as for the case of the inertial system discussed earlier and the system
extracts work provided $\frac{T_l}{T_h}<0.705$. Using same arguments similar to the underdamped case and keeping
 in mind only  the fact that there is only one  phase space variable, namely position, the average internal energy change 
in the overdamped limit  in the first step can be expressed as 
\begin{eqnarray}
 \Delta U_1 
=\frac{T_h}{2}-T_l.
\end{eqnarray}
Using the First law the average heat absorption from the hot bath during the first step is
\begin{equation}
 -Q_1=\Delta U_1 -\Delta F_h=\frac{T_h}{2}-T_l-\frac{T_h}{2}\ln\frac{1}{2}.
\label{o-q1}
\end{equation}
\begin{figure}
\vspace{0.5cm}

\includegraphics[width=7.5cm]{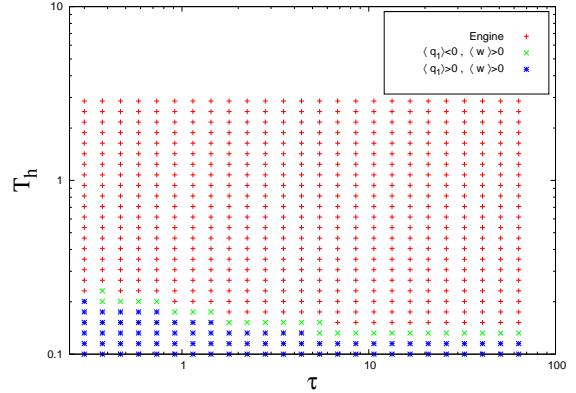}
\caption{(Color online) Phase diagram for different $T_h$ and $\tau$ but for fixed $T_l=0.1$.
}
\label{o-phase}
\end{figure}
The expression for  efficiency in the overdamped case is 
\begin{equation}
 \bar{\eta}_q=\frac{-W_{tot}}{-Q_1}=-\frac{\frac{T_h}{2}\ln\frac{1}{2}-\frac{T_h}{4}       +\frac{T_l}{2}\ln2+\frac{T_l}{2}}
                                {\frac{T_h}{2}-T_l-\frac{T_h}{2}\ln\frac{1}{2}},
\end{equation}
which is different from the earlier case. In quasistatic limit, from Eq.(\ref{o-q1}) heat flows from the bath to the system provided 
$ \frac{T_l}{T_h}< \frac{1+\ln 2}{2}=0.846$. This ratio $\frac{T_l}{T_h}$ differs from that obtained
for the underdamped case. From Eqs.(\ref{o-wtot}) and  (\ref{o-q1}), the system acts as an engine for the same condition ($\frac{T_l}{T_h}< 0.705$)
as for the underdamped case. A finite temperature difference between hot and cold bath is required so that the system can act as a heat engine.

\section{finite  cycle time engine in the Overdamped regime}
\label{onl}
\begin{figure}
\vspace{0.5cm}

\includegraphics[width=7.5cm]{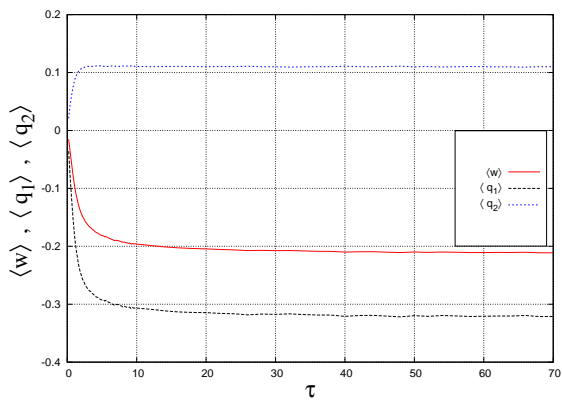}
\caption{(Color online) Variation of $\la w \ra$, $\la q_1\ra$ and $\la q_2\ra$ with  cycle time $\tau$.
}
\label{of-work}
\end{figure}

Analysis for finite time cycle is carried out by numerical methods as discussed earlier.
For a better understanding in the overdamped regime, all the  parameters have been kept same as  in the underdamped case.
In Fig.(\ref{o-phase}), we have plotted the phase diagram for the overdamped case keeping  $T_l$ fixed at 0.1.
For large $\tau$ ( quasistatic limit) we observe, from phase diagram, that the system operates as a heat engine provided
$T_h$ is greater than a critical value. This critical value is close to the theoretical value of 0.142
obtained from the bounds determined in quasistatic calculation. The phase diagram
shows a qualitative difference from the underdamped phase diagram (Fig.(\ref{uf-phase})).
The system always acts as an engine in $\tau\rightarrow0$ limit provided we are above a critical value of $T_h$,
which is not the case for the underdamped engine. This is clear from Fig.(\ref{of-work}), where we have plotted average work done 
on the system $\la w\ra$ and average heat released to each bath with $\la q_1\ra$ and $\la q_2\ra$ as a function 
of $\tau$. Note that the observed anomalous part for $\la w\ra$ and $\la q_1\ra$ in the underdamped case for small $\tau$ regime is absent in this
regime. The quantities $\la w\ra$, $\la q_1\ra$ and $\la q_2\ra$ show monotonic behavior and saturate at large cycle time to their
 analytical limits -0.214, -0.324 and 0.110, respectively. Unlike the underdamped case here, $\la w\ra$ and $\la q_1\ra$
are always negative.

\begin{figure}
\vspace{0.5cm}

\includegraphics[width=7.5cm]{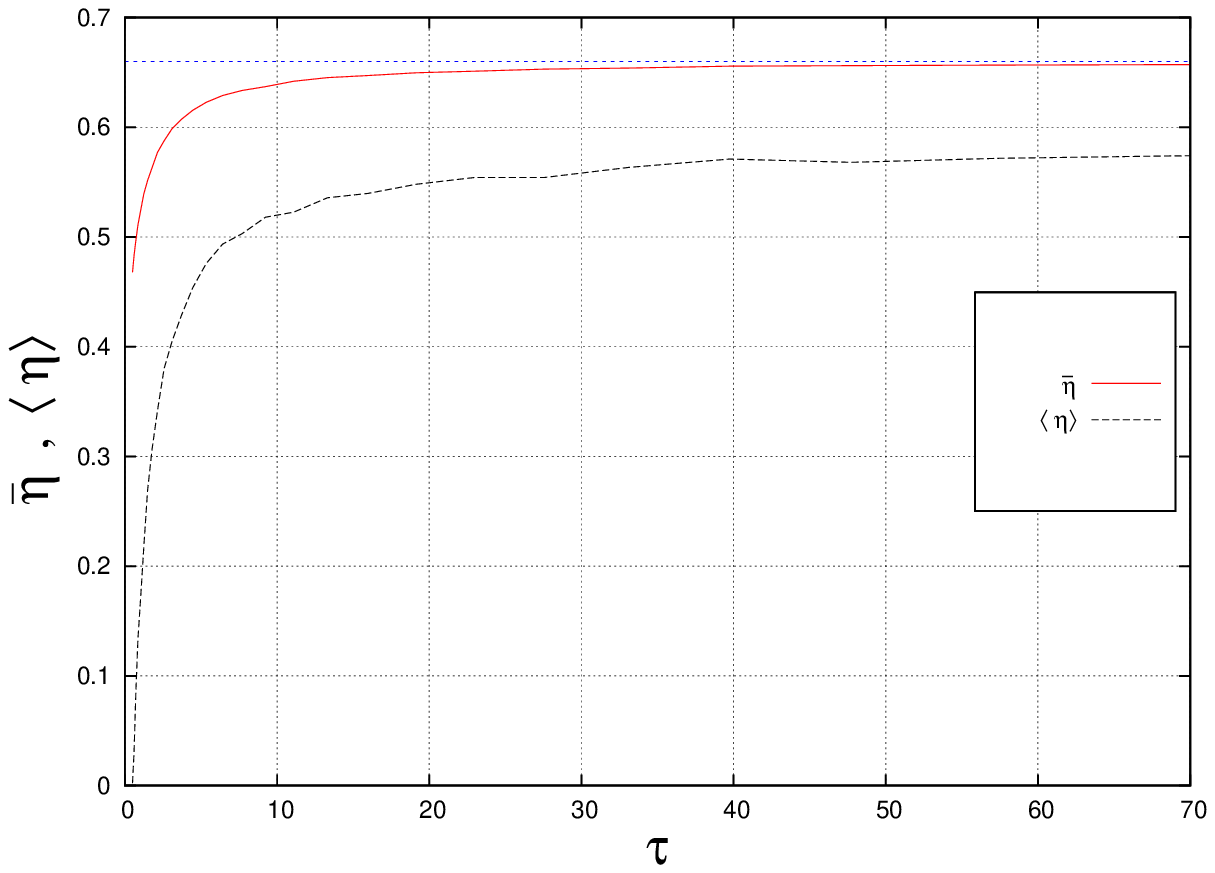}
\caption{(Color online) Variation of  $\la \eta\ra$  and $\bar{\eta}$ with  cycle time $\tau$. The doted blue line denotes the 
quasistatic limit for  $\bar{\eta}$.
}
\label{of-eff}
\end{figure}
\begin{figure}
\vspace{0.5cm}

\includegraphics[width=7.5cm]{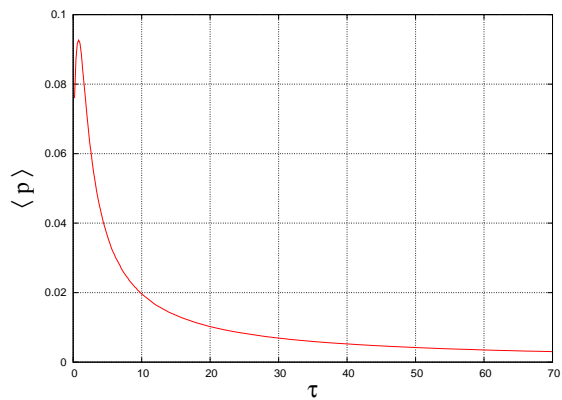}
\caption{(Color online) Variation of  $\la p\ra$ with  cycle time $\tau$.
}
\label{of-pow}
\end{figure}

In Fig.(\ref{of-eff}) we have plotted the average of efficiency $\la \eta\ra$ and $\bar{\eta}$ as a function of $\tau$. 
Both the efficiencies increase monotonically from zero and saturate for large $\tau$. The saturation value of 
$\bar{\eta}$ is close to the theoretically predicted value of 0.660. The saturation  value of $\la\eta\ra$ is found numerically
to be 0.571 which is much less than the corresponding value of $\bar{\eta}$. Both these values are less than the Carnot 
value $\eta_c=0.8$. It is clear that $\la\eta\ra \neq \bar{\eta}$ due to the strong correlation between 
fluctuating variables $w$ and $q_1$ for all  $\tau$.

From Fig.(\ref{of-pow}), we see that power exhibits a sharp peak at $\tau=0.8$. Corresponding efficiencies
$\la\eta\ra$ and $\bar{\eta}$ at maximum power are equal to 0.11 and 0.51 which are less than the C-A result ($\eta_{CA}=0.554$). 

\begin{figure}
\vspace{0.5cm}

\includegraphics[width=7.5cm]{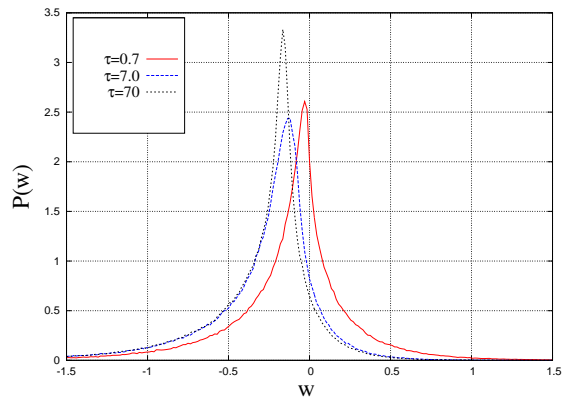}
\caption{(Color online) Distribution of $w$ for different cycle times in the overdamped case.
}
\label{ow-dist}
\end{figure}
\begin{figure}
\vspace{0.5cm}

\includegraphics[width=7.5cm]{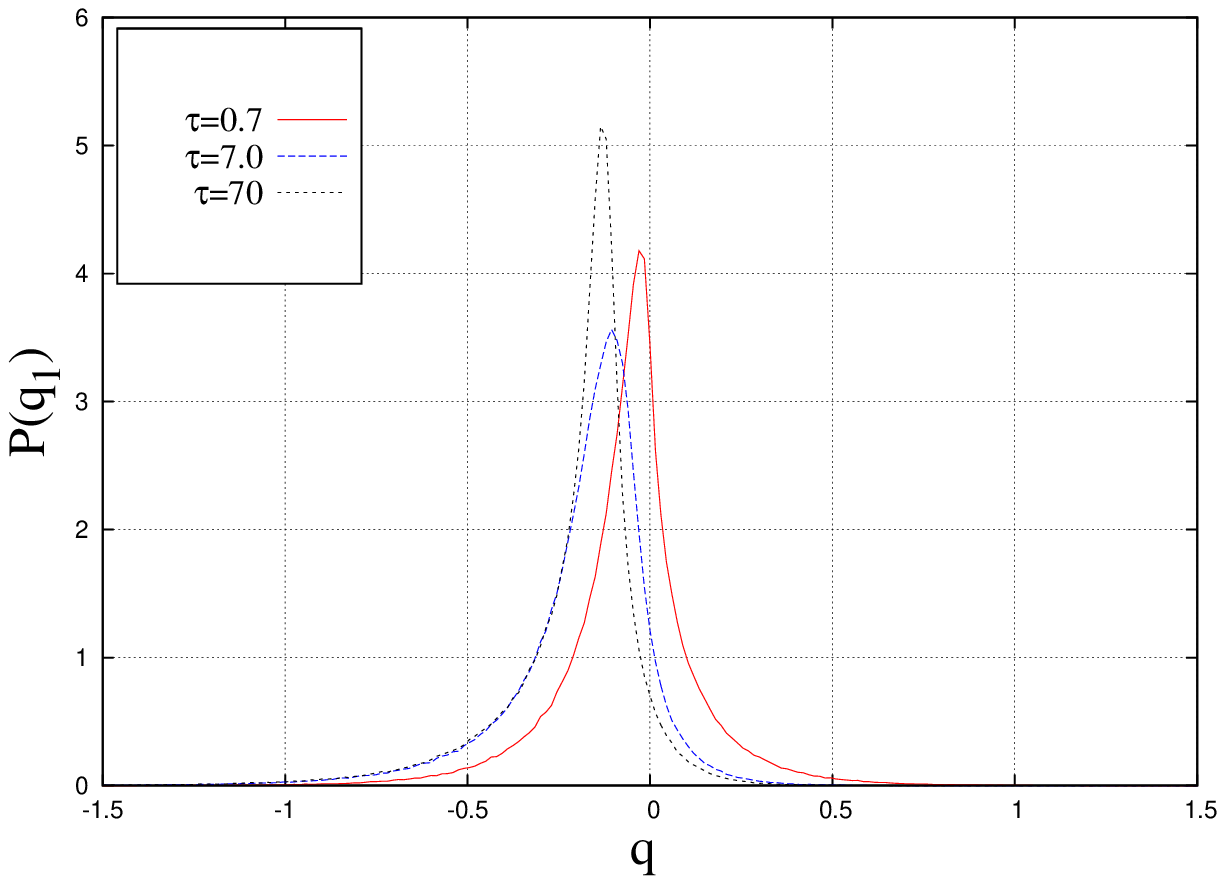}
\caption{(Color online) Distribution of $q_1$ for different cycle times in the overdamped case.
}
\label{oQ1-dist}
\end{figure}
\begin{figure}
\vspace{0.5cm}

\includegraphics[width=7.5cm]{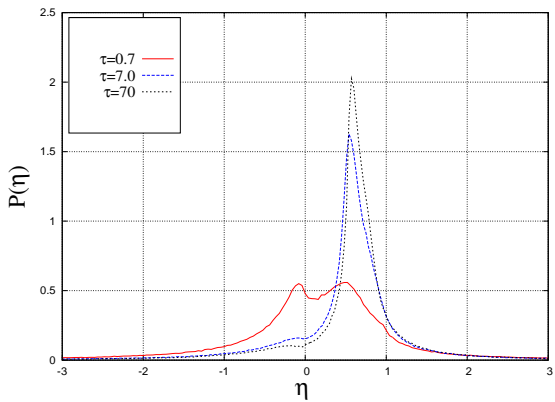}
\caption{(Color online) Distribution of $\eta$ for different cycle times in the overdamped case.
}
\label{oeff-dist}
\end{figure}
To study the nature of fluctuations in the overdamped regime we have plotted the distribution $P(w)$, $P(q_1)$
 and $P(\eta)$ in Figs.(\ref{ow-dist}), (\ref{oQ1-dist}) and (\ref{oeff-dist}) respectively. The qualitative
nature of the distributions of $P(w)$ and $P(q_1)$ remain the same for different values of $\tau$ as in the underdamped case.
The fluctuations are smaller compared to the underdamped case. The distribution $P(\eta)$   shows a double
peak behavior for $\tau=0.7$ with $\la \eta\ra=0.086$ and standard deviation $\sigma_{\eta}=1.688$. 
For  $\tau=7.0$, $\la \eta\ra=0.496$ and  $\sigma_{\eta}=1.287$.
For $\tau=70$, $\la \eta\ra=0.571$ and  $\sigma_{\eta}=1.234$. We observe that even in the
quasistatic regime, fluctuations of $\eta$ dominate over the mean value. Thus $\eta$ is a non self averaging quantity. We have also seen that the fraction
of the realizations for which the system acts as an engine, increases  with cycle time $\tau$. Numerical values for these 
fractions are 0.488, 0.817 and 0.861, for $\tau=0.7$, 7.0 and 70.0, respectively. Hence, finite fraction of realization 
does not act as an engine even in quasistatic limit.  Similar to the underdamped case, 
the reliability of the system to act as an engine increases with $\tau$.

Finally, we discuss the performance characteristics of our system in the overdamped regime using an experimental protocol 
 \cite{nphy12}. The experimental protocol consists of only two steps, in which the two adiabatic steps of Fig.(\ref{fig-pro})
are absent. In the quasistatic regime the system acts as an engine for any temperature difference and there is no bound
on $T_h$ unlike our four step protocol \cite{nphy12}. This  suggests that the phase diagram will depend on the nature 
of protocol as well as on system parameters and is not unique. As discussed earlier, most of the work fluctuations
specially in quasistatic regime  arise from two adiabatic 
steps. In the absence of these two steps, we have observed in our simulation that in the quasistatic regime, work distribution $P(w)$ is sharply 
peaked like a delta function at  $W=-\frac{1}{2}(T_h-T_l)\ln 2$ (analytical result) \cite{nphy12}. However, fluctuations in $q_1$
persist even in the quasistatic regime as a result of the relaxation process that follows when the system, in contact
 with the cold bath, is brought in direct contact with high temperature reservoir. The distribution of stochastic efficiency
$P(\eta)$ exhibits a qualitative differences. It has almost zero weight for $\eta<0$ in large $\tau$ limit and shows
a broad double peak feature which is confined in the region $0<\eta<1$. Beyond $\eta>1$ a long tail is observed.
For $\tau=70$ we have numerically calculated  $\la\eta\ra=0.579$ and $\sigma_{\eta}=0.903$. Even for this 
protocol we notice that fluctuations dominates over the mean value. The details of these results will be published elsewhere 
\cite{shu00}.

\section{Summary}
\label{con}
We summarise our results in this section. We have  carried out an extensive analysis of a single particle stochastic heat engine 
by manipulating a Brownian particle in a harmonic trap with a periodically time dependent stiffness constant as a protocol.
The cycle consists of two isothermal steps and two adiabatic steps similar to that of Carnot engine. The proposed model
is studied taking into account both the inertial and overdamped Langevin equations. Thermodynamic quantities, defined over
 microscopic phase space trajectory of our system, fluctuate from one cycle
 of operation to another. Their magnitude depends on the trajectory of the particle during the cycle.
This is done by using the methods of stochastic energetics. Average value of thermodynamic quantities and their 
distribution functions have been calculated numerically in TPSS. Analytical results of average thermodynamic 
quantities have been obtained in the quasistatic regime. These results are consistent with the corresponding numerical
results. We have reported several new results which were not addressed in earlier literature.

 The full phase diagram for operation of a system is given in both inertial and high friction regime.
They differ from each other qualitatively. In both cases it is also shown that system acts as an engine
provided the temperature difference is greater than a critical value (unlike Carnot engine). This 
critical value depends on system parameters and is consistent with analytical results  in quasistatic limit.
Moreover, for fixed bath temperatures and system parameters there should be a critical cycle time 
above which the system acts as an engine.

  The mean of the stochastic efficiency is dominated by its fluctuations ($\la\eta\ra<\sigma_{\eta}$)
 even in quasistatic regime, making the efficiency a non-self averaging quantity.
 For such cases mean ceases to be a good physical
 variable and one has to resort to the analysis for full probability distribution. This is one of our main result.
 We have also shown that $\bar{\eta}=\frac{\la w\ra}{\la q_1\ra}\ne\la\frac{w}{q_1}\ra=\la\eta\ra$.

 Our analysis of  model dependence of finite cycle time clearly rules out any simple universal relation 
( e.g., $\eta_{CA}=1-\sqrt\frac{T_l}{T_h}.$) between efficiency at maximum power and temperature of the baths.
 Time asymmetric periodic protocol makes engine less efficient. Only in the quasistatic regime time asymmetry does not 
play any role.

 For given cycle time there are several realizations which do not work as a heat engine. These are referred to as 
transient second law violating trajectories. Number of these realizations decreases 
as we increase $\tau$. The fractions of realisations following second law with corresponding $\tau$ are reported earlier sections both in underdamped and overdamped regimes. Thus for large cycle time the reliability of the system working as an engine increases.
Persistence of these realizations even in quasistatic regime can be attributed to the fluctuation of heat and work distributions. Fluctuations in work are mainly attributed to two adiabatic processes connecting two isotherms, while fluctuations of $q_1$ result from the relaxation of the system, when brought in direct contact with high temperature reservoir from low temperature  bath.
.

We have shown that in TPSS $P_{ss}(x,v,t)$ exhibit strong correlation between variables $x$ and $v$ in small cycle time limit.
However, it becomes uncorrelated as we approach  quasistatic limit. For analytical simplicity it had been generally assumed in earlier literature that there is no correlation 
between $x$ and $v$ in  $P_{ss}(x,v,t)$ (see for example \cite{tu13}).

 In the inertial regime we have also verified the recently proposed fluctuation theorems for heat engines in a TPSS. 
  Our results are amenable to experimental verifications.

\section{ACKNOWLEDGMENTS}
A.M.J. thanks DST, India for financial support and. A.S. thanks MPIPKS, Germany for partial support.

\end{document}